\documentclass[11pt]{article}
\usepackage{graphicx}\usepackage{amsmath,amsthm,amssymb}

\begin{document}

\title{\textbf{\textsf{Primordial Black Holes in Phantom
Cosmology}}}

\author{\textsf{ Mubasher Jamil}\footnote{mjamil@camp.nust.edu.pk}\ \ \textsf{and Asghar Qadir}\footnote{aqadirmath@yahoo.com}\\
 \\
\textit{Center for Advanced Mathematics and Physics}\\
\textit{ National University of Sciences and Technology}\\
\textit{Campus of College of E\&ME}\\
\textit{Peshawar Road, Rawalpindi, 46000, Pakistan} \\
} \maketitle

\begin{abstract}

We investigate the effects of accretion of phantom energy onto
primordial black holes. Since Hawking radiation and phantom energy
accretion contribute to a {\it decrease} of the mass of the black
hole, the primordial black hole that would be expected to decay now
due to the Hawking process would decay {\it earlier} due to the
inclusion of the phantom energy. Equivalently, to have the
primordial black hole decay now it would have to be more massive
initially. We find that the effect of the phantom energy is
substantial and the black holes decaying now would be {\it much}
more massive --- over 10 orders of magnitude! This effect will be
relevant for determining the time of production and hence the number
of evaporating black holes expected in a universe accelerating due
to phantom energy.

\end{abstract}


\indent \large
\newpage

\section{Introduction}

Numerous astrophysical observations are consistent with the standard
cold dark matter model with the inclusion of an effective
cosmological constant. A classical cosmological constant is
generally avoided as quantum gravity attempts lead to a natural
expectation of a cosmological term of Planck scale, which is totally
at odds with the value required by observation (see for example
\cite{ls}). As such, it is generally assumed that there is some
physical field that comes into play {\it after the Planck era} and
there is some principle that excludes the induced Planck energy
cosmological term (see for example \cite{gth}). This exotic field is
often called ``dark energy". According to the generally accepted
modeling, the latter constitutes more than 70$\%$ of the total
energy density of the universe while the matter component carries
most of the remaining part \cite{perl,ries,sper,sper1,cope}. It is
not clear how seriously to take the ``quantum gravity" requirements
considering that there is no viable theory of quantum gravity to
date \cite{ls,rp}. Further, there is no clear evidence that the
onset of quantum gravity will be at Planck scale and not orders of
magnitude away from it \cite{aq}. There are alternate (classical)
explanations of the observations available in the literature (for
example by Wiltshire \cite{dw}) but here we shall follow the
generally accepted view of some form of dark energy providing the
observed acceleration of the Universe.

The observable universe locally appears to be spatially flat with an
equation of state (EoS) parameter (the ratio of pressure to the
energy density) $\omega (\equiv p_x/\rho_x)\simeq-1.$ The dark
energy is an exotic vacuum energy with negative pressure and
positive energy density which arises due to quantum vacuum
fluctuations in spacetime. Caldwell and co-workers
\cite{caldwell,caldwell1} considered the possibility of dark energy
with the super-negative EoS parameter $\omega<-1$, which they called
`phantom energy'. It also gives negative pressure. The motivation to
consider phantom energy as the candidate for dark energy arises from
the observational data of the cosmic microwave background power
spectrum and supernovae of type Ia. The phantom energy violates the
general relativistic energy conditions including the null and
dominant ones \cite{qiu}. Its implications in cosmology give rise to
exotic phenomena like an imaginary value of the sound speed,
negative temperature, the divergence of the scale factor $a(t)$ and
the energy density $\rho_x\sim a^{-3(1+\omega)}$, at a finite time
resulting in a `big rip', an epoch when the spacetime is torn apart
\cite{pedro11,fabris,mar,vinod,carroll} (see \cite{nojiri} for a
review on the big rip singularity). However there are some attempts
made recently in which the occurrence of a big rip singularity is
avoided by phantom energy decay into matter
\cite{anton,carroll1,pedro}. Another attempt is the `big trip', in
which a wormhole accretes phantom energy and grows so large that it
engulfs the whole universe \cite{moruno}. A similar scenario is also
proposed for black holes whereby the black hole event horizon
inflates to swallow up the cosmological horizon, resulting in a
naked singularity \cite{martin11}. Moreover, quantum gravitational
effects (if they exist) may avoid the big rip singularity. If the
big rip cannot be avoided, the smaller the parameter $\omega$, the
closer the big rip will be to the present time.

Another weird property of phantom energy is that its accretion onto
gravitationally bound structures results in their dissociation and
disintegration in a rather slow process. It was first analyzed in
\cite{ness} for several gravitational systems like the solar system
and the Milky Way galaxy. Initially, this possibility was
investigated for a Schwarzschild black hole by Babichev et al
\cite{babichev}, who showed that the black hole mass goes to zero
near the big rip. Interestingly, in this scenario larger black holes
lose mass more rapidly than smaller ones. Later on, their model was
extended to the Reissner-Nordstrom \cite{jamil}, Schwarzschild de
Sitter \cite{martin} and Kerr-Newmann \cite{babichev1,jose} black
holes. It should be mentioned that it has been argued that the
mechanism of accretion followed by Babichev et al is stationary and
does not possess the shift symmetry \cite{vikman} and hence that the
mechanism of dark energy accretion is not realistic and consistent.
Nevertheless, we shall follow the Babichev et al analysis, leaving
the detailed analysis for subsequent work, as the effect will be
technically difficult to apply and we believe will not make a
substantial difference for phantom energy in the neighbourhood of a
primordial black hole. It has also been argued that when the
back-reaction effects of the accretion process are included in the
analysis of Babichev et al \cite{babichev}, the black hole mass may
increase instead of decreasing \cite{nojiri1,gao}, thus avoiding the
big rip. Also in cyclic cosmological models, black holes do not tear
apart near the turnaround but preserve some nonzero mass
\cite{sun,zhang}. We shall ignore the big rip issue here.

\section{Hawking evaporation of black holes}

We are interested in studying the effects of accretion of phantom
energy on a static primordial black hole. Carr and Hawking
\cite{haw} in 1974 considered the formation of black holes of mass
$10^2$kg and upwards in the early evolution of the universe. After
their attempt, several authors investigated various scenarios of PBH
formation \cite{khlopov,barrow1,raf,polarski}. The existence of
these small mass black holes was based on the assumption that the
early universe was not entirely spatially smooth but there were
density fluctuations or inhomogeneities in the primordial plasma
which gravitationally collapsed to form these black holes. Unlike
the conventional black holes that are formed by the gravitational
collapse of stars or mergers of neutron stars, the primordial black
holes (PBH) are formed due to the gravitational collapse of matter
without forming any initial stellar object. The mass of a PBH can be
of the order of the particle horizon mass at the time of its
formation \cite{carr,carr1}
\begin{equation}
M_{PBH}\approx\frac{c^3t}{G}\approx10^{12}\left(\frac{t}{10^{-23}s}
\right)kg~.
\end{equation}
Therefore PBHs that formed in the early history of the universe must
be less massive while those that formed later must be more massive.
Black holes formed at Planck time $10^{-43}$s would have Planck mass
$10^{-8}$kg.

Using classical arguments, Penrose and Floyd showed that one can
extract rotational energy from a rotating black hole \cite{pf}.
Penrose went on to argue (see \cite{rp} and references therein) that
one could take thermal energy from the environs of a black hole and
throw it into the black hole to get usable energy out. This would
apparently reduce the entropy around the black hole. As such, he had
argued that {\it there must be an entropy of the black hole that
increases at least as much as that of its environs decreases}.
Hawking had pointed out that in any physical process the area of a
black hole always increases \cite{swh} just as entropy always
increases. This led Bekenstein \cite{jacob} to propose a linear
relationship between the area and entropy of a black hole. Thus
Bekenstein \cite{jacob1,jacob2} generalized the second law of
thermodynamics to state that the sum of the entropy of the black
hole and its environs never decreases. However, at this stage it
seemed that the connection between black holes and thermodynamics
was purely formal. At this stage Fulling pointed out that
quantization of scalar fields in accelerated frames gives an
ambiguous result \cite{saf}, which seemed to yield radiation seen in
the accelerated with a fractional number of particles. Hawking
repeated the calculation for an observer near a black hole and
obtained the same result by various methods and found that the
radiation had a thermal spectrum \cite{hawking}. This led him to
propose that mini-PBHs would evaporate away in a finite time
\cite{hawk}.

The corresponding Hawking evaporation process reduces the mass of
the black hole by \cite{page}
\begin{equation}
\left.\frac{dM}{dt}\right\vert_{hr}=-\frac{\hbar
c^4}{G^2}\frac{\alpha}{M^2},
\end{equation}
where $\alpha$ is the spin parameter of the emitting particles.
Integration of Eq. (2) gives the evolution of PBH mass as
\begin{equation}
M_{hr}=M_i\left(1-\frac{t}{t_{hr}}\right)^{1/3},
\end{equation}
where the Hawking evaporation time scale $t_{hr}$ is
\begin{equation}
t_{hr}=\frac{G^2}{\hbar c^4}\frac{M_i^3}{3\alpha}.
\end{equation}
It is obvious from Eq. (3) that as $t\rightarrow t_{hr}$, the mass
$M_{hr}\rightarrow0$. Plugging in $t_{hr}=t_o$ (the current age of
the universe) in Eq. (4) gives the mass $10^{12}\text{kg}$ of the
PBH that should have been evaporating now. Hence from Eq. (1), it
can be estimated that these PBHs were formed before about
$10^{-23}\text{sec}$. For $M_i\gg10^{14}kg,$
$\alpha=2.011\times10^{-4}$, hence Eq. (4) implies
$t_{hr}\simeq2.16\times10^{-18}\left(\frac{M}{kg}\right)^3\text{sec}$.
While for $5\times10^{11}kg\ll M_i\ll10^{14}\text{kg}$,
$\alpha=3.6\times10^{-4}$ then Eq. (4) gives
$t_{hr}\simeq4.8\times10^{-18}\left(\frac{M}{kg}\right)^3\text{sec}$.
Therefore detecting PBHs would be a good tool to probe the very
early universe (closer to the Planck time). The evaporation of PBHs
could still have interesting cosmological implications: they might
generate the microwave background \cite{zel} or modify the standard
cosmological nucleosynthesis scenario \cite{novikov} or contribute
to the cosmic baryon asymmetry \cite{barrow}. Some authors have also
considered the possibility of the accretion of matter and dust onto
the seed PBH resulting in the formation of super-massive black holes
which reside in the centers of giant spiral and elliptical galaxies
\cite{bean}.

\section{Phantom energy accretion onto black hole}

The FRW equations governing the dynamics of our gravitational system
are given by
\begin{eqnarray}
H^2\equiv\left(\frac{\dot{a}}{a}\right)^2&=&\frac{8\pi G}{3}(\rho_m+\rho_x),\\
\frac{\ddot{a}}{a}&=&-\frac{4\pi G}{3}[\rho_m+\rho_x(1+\omega)]~.
\end{eqnarray}
Here $\rho_m$ and $\rho_x$ denote the energy densities of matter and
the exotic energy densities respectively. The scale factor $a(t)$
goes like \cite{vinod}
\begin{equation}
a(t)=\frac{a(t_0)}{[-\omega+(1+\omega)t/t_0]^{-\frac{2}{3(1+\omega)}}}
\ \ (t>t_0),
\end{equation}
where $t_0$ is the time when the universe transits from matter to
exotic energy domination (which is roughly equal to the age of the
universe). Notice that the scale factor $a(t)$ diverges when the
quantity in the square brackets in Eq.(7) vanishes identically i.e.
\begin{equation}
t^*=\frac{\omega}{1+\omega}t_0~.
\end{equation}
Subtracting $t_0$ from Eq. (8), we get
\begin{equation}
t^*-t_0=\frac{1}{1+\omega}t_0~.
\end{equation}
The evolution of energy density of the exotic energy is given by
\begin{equation}
\rho_{x}^{-1}=6\pi G(1+\omega)^2(t^*-t)^2.
\end{equation}
A black hole accreting only the exotic energy has the following rate
of change in mass \cite{babichev}
\begin{equation}
\left.\frac{dM(t)}{dt}\right\vert_x=\frac{16\pi G^2}{c^5}
M^2(\rho_{x}+p_{x})~.
\end{equation}
It is clear that when $\rho_x+p_x<0$, the mass of the black hole
will decrease. We are particularly interested in the evolution of
black holes about and after $t=t_0$ since the dark energy is
presumably negligible before that time and may not have any
noticeable effects on the black hole. Using Eqs. (9) and (10) in
(11), we get
\begin{equation}
\left. \frac{dM(t)}{dt}\right\vert
_{x}=\frac{8G}{3c^3}\frac{M^2}{t_0^2}(1+\omega)~.
\end{equation}
Therefore the mass change rate for a black hole accreting pure
exotic energy is determined by Eq. (12). For the phantom energy
accretion, the time scale is obtained by integrating Eq. (12) to get
\begin{equation}
M(t)=M_i\left( 1-\frac{t}{t_x} \right)^{-1},
\end{equation}
where $t_x$ is the characteristic accretion time scale given by
\begin{equation}
t_x^{-1}=\frac{16\pi G^2}{c^5} M_i(\rho_x+p_x)~.
\end{equation}
Using Eqs. (9) and (10) in (14), we get
\begin{equation}
t_x=\frac{3c^3}{8G}\frac{t_0^2}{M_i(1+\omega)}~.
\end{equation}

\section{Evolution of mass due to phantom energy accretion and Hawking evaporation}

The expression determining the cumulative evolution of the black
hole is obtained by adding Eqs. (2) and (12) i.e.
\begin{eqnarray}
\left.\frac{dM(t)}{dt}\right\vert_{Total}&=&\left.\frac{dM}{dt}\right\vert_{hr}+\left.\frac{dM}{dt}\right\vert_{x},\\
&=& -\frac{\hbar
c^4}{G^2}\frac{\alpha}{M^2}+\frac{8G}{3c^3}\frac{M^2}{t_0^2}(1+\omega).
\end{eqnarray}
We write the above equation as
\begin{equation}
\frac{dM}{dt}=-aM^2-\frac{b}{M^2},
\end{equation}
where
\begin{equation}
a=\frac{8G}{3c^3}\frac{\epsilon}{t_0^2},\ \ b=\frac{\hbar
c^4\alpha}{G^2}.
\end{equation}
Here $\epsilon=-\omega-1$. Thus (18) can be written in the form
\begin{equation}
-\int dt=\frac{1}{b}\int\frac{M^2dM}{1+\frac{a}{b}M^4}\nonumber
\end{equation}
To integrate above equation, we assume
\begin{equation}
x=\Big(\frac{a}{b}\Big)^{1/4}M,
\end{equation}
which yields
\begin{equation}
-\int dt=\frac{1}{(a^3b)^{1/4}}\int\frac{x^2dx}{1+x^4},
\end{equation}
We note that \cite{book}
\begin{eqnarray}
\int\frac{x^{m-1}dx}{1+x^{2n}}&=&-\frac{1}{2n}\sum_{k=1}^{n}\cos
\Big(\frac{m\pi (2k-1)}{2n}\Big)\ln \Big| 1-2x\text{cos}
\Big(\frac{2k-1}{2n}\Big)\pi+x^2 \Big|\nonumber
\\&\;&+\frac{1}{n}\sum_{k=1}^{n}\sin \Big(\frac{m\pi
(2k-1)}{2n}\Big)\tan^{-1}\Big[\frac{x-\cos
\Big(\frac{2k-1}{2n}\Big)\pi}{\sin
\Big(\frac{2k-1}{2n}\Big)\pi}\Big],\ \ m<2n.
\end{eqnarray}
In our case, $m=3$ and $n=2$, hence the above equation yields
\begin{equation}
\int\frac{x^2dx}{1+x^4}=\frac{1}{4\sqrt{2}}\ln \Big|
\frac{1-\sqrt{2}x+x^2}{1+\sqrt{2}x+x^2}
\Big|+\frac{1}{2\sqrt{2}}\tan^{-1}\Big( \frac{\sqrt{2}x}{1+x^2}
\Big).
\end{equation}
On substituting the value of $x$ above, we obtain
\begin{equation}
t=t_0+\frac{1}{4\sqrt{2}}\ln\Big|\frac{1-\sqrt{2}\Big(\frac{a}{b}\Big)^{1/4}M+
\Big(\frac{a}{b}\Big)^{1/2}M^2}{1+\sqrt{2}\Big(\frac{a}{b}\Big)^{1/4}M+\Big(\frac{a}{b}\Big)^{1/2}M^2}
\Big|+\frac{1}{2\sqrt{2}}\tan^{-1}\Big[
\frac{\sqrt{2}\Big(\frac{a}{b}\Big)^{1/4}M}{1+\Big(\frac{a}{b}\Big)^{1/2}M^2}
\Big].
\end{equation}
We now redefine the values of $a$ and $b$ by assuming $M=mM_i$,
where $m$ is a dimensionless parameter and $M_i$ is the initial mass
of the black hole. Thus (18) becomes
\begin{equation}
\frac{dm}{dt}=-a^\prime m^2-\frac{b^\prime}{m^2},
\end{equation}
where $a^\prime=aM_i$ and $b^\prime=b/M_i^3$. For the terms to be
equal strength, we require $a^\prime\approx b^\prime$. Thus
\begin{equation}
M_i\approx\Big( \frac{b}{a} \Big)^{1/4}.
\end{equation}
Now
\begin{equation}
\frac{b}{a}=\frac{3\hbar c^7t_0^2\alpha}{8G^3\epsilon},\ \
\text{or},\ \ \epsilon=\frac{3\hbar c^7t_0^2\alpha}{8G^3M_i^4}.
\end{equation}
We can normalize
\begin{equation}
t=t_0\left[1-\frac{\frac{1}{4\sqrt{2}}\ln\Big|\frac{1-\sqrt{2}\Big(\frac{a^\prime}
{b^\prime}\Big)^{1/4}m+\Big(\frac{a^\prime}{b^\prime}\Big)^{1/2}m^2}{1+\sqrt{2}\Big(\frac{a^\prime}{b^\prime}\Big)^{1/4}m+\Big(\frac{a^\prime}{b^\prime}\Big)^{1/2}m^2}
\Big|+\frac{1}{2\sqrt{2}}\tan^{-1}\Big[
\frac{\sqrt{2}\Big(\frac{a^\prime}{b^\prime}\Big)^{1/4}m}{1+\Big(\frac{a^\prime}{b^\prime}\Big)^{1/2}m^2}
\Big]}{\frac{1}{4\sqrt{2}}\ln\Big|\frac{1-\sqrt{2}\Big(\frac{a^\prime}{b^\prime}\Big)^{1/4}+
\Big(\frac{a^\prime}{b^\prime}\Big)^{1/2}}{1+\sqrt{2}\Big(\frac{a^\prime}{b^\prime}\Big)^{1/4}+
\Big(\frac{a^\prime}{b^\prime}\Big)^{1/2}}
\Big|+\frac{1}{2\sqrt{2}}\tan^{-1}\Big[
\frac{\sqrt{2}\Big(\frac{a^\prime}{b^\prime}\Big)^{1/4}}{1+\Big(\frac{a^\prime}{b^\prime}\Big)^{1/2}}
\Big]}\right].
\end{equation}
Replacing $p=a^\prime/b^\prime=\frac{8\epsilon G^3}{3\alpha\hbar
c^7t_0^2}M_i^4 \sim M_i^4$ (the ratio of the phantom component to
the Hawking component, in the energy radiated) the above equation
becomes
\begin{equation}
t=t_0\left[1-\frac{\frac{1}{4\sqrt{2}}\ln\Big|\frac{1-\sqrt{2}p^{1/4}m+p^{1/2}m^2}
{1+\sqrt{2}p^{1/4}m+p^{1/2}m^2}
\Big|+\frac{1}{2\sqrt{2}}\tan^{-1}\Big(
\frac{\sqrt{2}p^{1/4}m}{1+p^{1/2}m^2}
\Big)}{\frac{1}{4\sqrt{2}}\ln\Big|\frac{1-\sqrt{2}p^{1/4}+p^{1/2}}{1+\sqrt{2}p^{1/4}+p^{1/2}}
\Big|+\frac{1}{2\sqrt{2}}\tan^{-1}\Big(
\frac{\sqrt{2}p^{1/4}}{1+p^{1/2}} \Big)}\right].
\end{equation}
Moveover, the power emission due to Hawking evaporation from the
stationary black hole of mass $M\gg10^{17}\text{g}$ \cite{page}
\begin{equation}
P=3.458\times10^{46}(M/\text{g})^{-2}\text{erg}\text{s}^{-1},
\end{equation}
and for mass $5\times 10^{14}\text{g}\ll M\ll10^{17}\text{g}$,
\begin{equation}
P\approx3.6\times10^{16}(M/10^{15}\text{g})^{-2}\text{erg}\text{s}^{-1}.
\end{equation}
In our analysis, the mass in the above two expressions is replaced
by
\begin{equation}
M=\Big( \frac{3\hbar c^7t_0^2\alpha}{8G^3\epsilon}
\Big)^{1/4}\text{g}.
\end{equation}
Now choosing $\epsilon=0.1$, we obtain $M=8.74029\times10^{22}$g
which will be evaporating now due to the combined effects of phantom
energy and Hawking radiation. Then using Eq. (30), the corresponding
power emission will be $P=4.52661~\text{erg}$  $\text{s}^{-1}$. We
can compare this result with that of a black hole of mass
$M\simeq1.05\times10^{12}\text{g}$ evaporating just now due to
Hawking radiation only. The corresponding power emission will be
$P\simeq3.144\times10^{22}\text{erg}$ $\text{s}^{-1}$. Note that the
power emission from a black hole decreases when the effects of
phantom energy are incorporated. Similarly, for very large values of
$\epsilon \sim 10^{25}$ would give
$M=2.763923\times10^{16}\text{g}$. Using this mass in (31), the
power emitted is $4.52661\times10^{13}\text{erg}$ $\text{s}^{-1}$.
However, such large values would lead to a very early big rip and
hence must be excluded. Thus black holes $\sim10^{22}\text{g}$ are
of more interest for observational purposes since these are the ones
that should be evaporating now.

\section{Conclusion}

In this paper, we have analyzed the Hawking radiation effects
combined with the phantom energy accretion on a stationary black
hole. The former process has been thoroughly investigated in the
literature. However there is as yet no observational support to it.
According to standard theory it is assumed that after the formation
of PBHs (of mass $\sim10^{12}\text{kg}$ with a Hawking temperature
$10^{12}\text{K}$), they would absorb virtually no radiation or
matter whatsoever during their evolution and radiate continuously
till they evaporate in a burst of gamma rays at the present time.
This scenario assumes that the Hawking temperature for such black
holes was always larger than the background temperature of the CMB.
Strictly speaking, this cannot be true. Consequently PBHs could have
accreted the background radiation (and even some matter) and {\it
grown} in mass. Hence there should be no PBH left to be evaporating
right now \cite{siva}. However, the above scenario is modified when
phantom energy comes into play. When phantom energy and the Hawking
process are relevant the total life time scale of the PBH is
significantly shortened and the formation of the PBH exploding now
is delayed.

From Eq. (29) we obtain the time as a function of mass instead of
getting mass as a function of time. To make sense of the results we
need to obtain the evolution with time. This is done by inverting
the explicit function. We have plotted the normalized time
$\tau=t/t_0$ against the dimensionless mass parameter $m$ and $m$
against $\tau$ for different choices of the parameter $p$, in
Figures 1 - 10. It is observed that increasing $p$ increases the
steepness of the curve specifying the mass evolution. Therefore the
black hole loses mass faster for larger $p$ till it vanishes at
$\tau=1$, the present time. In particular, Figures 7 and 9 show the
same evolution of mass for larger values of $p$. It appears that the
graphs contain a redundant (or nonphysical) part of the mass
evolution and the only physically interesting section is above the
horizontal curve crossing $t=0$. Thus in effect, see Figures 8 and
10, the initial mass of the black hole must be taken $0.45M_i$ of
the value given by for $p=5$ and about $0.315M_i$ for $p=10$. It is
obvious that the results are very insensitive to changes of the
parameter $\epsilon$ for the phantom energy. As such, they can be
regarded as fairly robust.

\subsubsection*{Acknowledgment}
One of us (MJ) would like to the thank the National University of
Sciences and Technology for full support for participation in the
Second Italian-Pakistan Workshop on Relativistic Astrophysics. One
of us (AQ) would like to thank ICRANet and Prof. Remo Ruffini for
support and hospitality at Pescara, Italy.

\newpage
\begin{figure}
\includegraphics{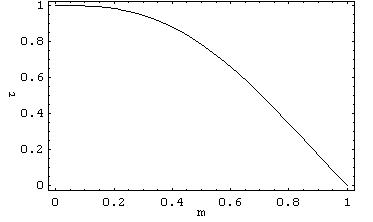}\\
\caption{The normalized time is plotted against the mass
parameter for $p=0.1$.}
\end{figure}
\newpage
\begin{figure}
\includegraphics{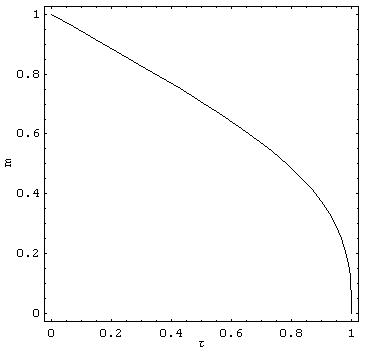}\\
\caption{The mass parameter is plotted against the normalized time
for $p=0.1$.}
\end{figure}
\newpage
\begin{figure}
\includegraphics{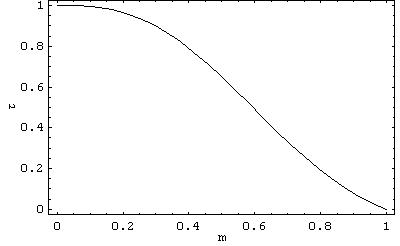}\\
\caption{The normalized time is plotted against the mass parameter
for $p=0.5$.}
\end{figure}
\newpage
\begin{figure}
\includegraphics{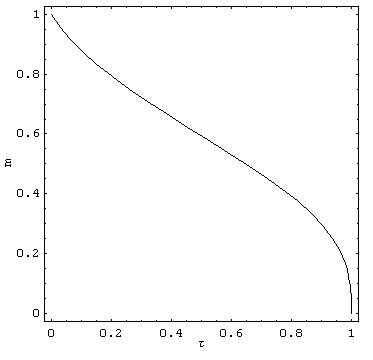}\\
\caption{The mass parameter is plotted against the normalized time
for $p=0.5$.}
\end{figure}
\newpage
\begin{figure}
\includegraphics{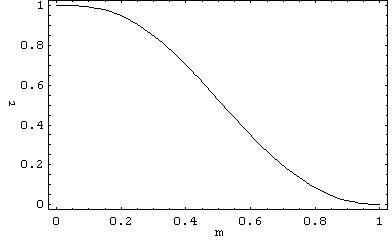}\\
\caption{The normalized time is plotted against the mass parameter
for $p=1$.}
\end{figure}
\newpage
\begin{figure}
\includegraphics{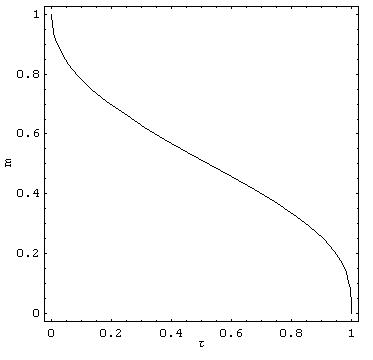}\\
\caption{The mass parameter is plotted against the normalized time
for $p=1$.}
\end{figure}
\newpage
\begin{figure}
\includegraphics{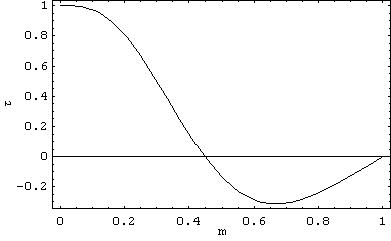}\\
\caption{The normalized time is plotted against the mass parameter
for $p=5$.}
\end{figure}
\newpage
\begin{figure}
\includegraphics{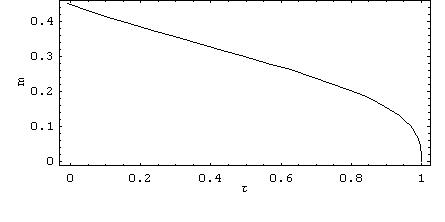}\\
\caption{The mass parameter is plotted against the normalized time
for $p=5$.}
\end{figure}
\newpage
\begin{figure}
\includegraphics{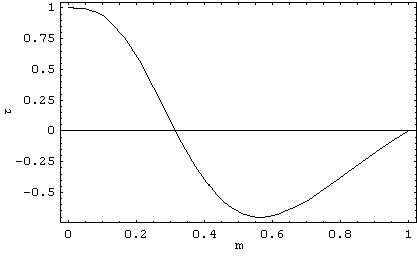}\\
\caption{The normalized time is plotted against the mass parameter
for $p=10$.}
\end{figure}
\newpage
\begin{figure}
\includegraphics{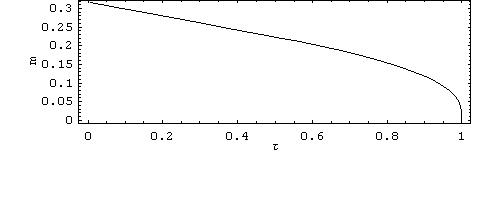}\\
\caption{The mass parameter is plotted against the normalized time
for $p=10$.}
\end{figure}


\newpage


\begin{thebibliography}{99}
\bibitem{ls} L. Smolin, {\it The Trouble With Physics: The Rise
of String Theory, the Fall of a Science, and What Comes Next},
Houghton Mifflin 2006.
\bibitem{gth} G. 't Hooft, {\it Proceedings of the 12th Regional
Conference on Mathematical Physics}, eds. M.J. Aslam, F. Hussain, A.
Qadir, Riazuddin and H. Saleem, World Scientific, 2007.
\bibitem{perl} S. Perlmutter et al, \textit{Ap. J}. 517 (1999) 565.
\bibitem{ries} A.G. Riess et al,\textit{ Astron. J.} 116 (1998) 1009.
\bibitem{sper} D.N. Spergel et al, \textit{Ap. J.} 148 (2003) 175.
\bibitem{sper1} D.N. Spergel et al, \textit{Ap. J. }170 (2007) 377.
\bibitem{cope} E.J. Copeland et al,
\textit{Int.J.Mod.Phys.}D15 (2006) 1753.
\bibitem{rp} R. Penrose, {\it The Road to Reality: A Complete Guide
to the Laws of the Universe}, Vintage Books 2004.
\bibitem{aq} A. Qadir, {\it Proceedings of the 12th Regional
Conference on Mathematical Physics}, eds. M.J. Aslam, F. Hussain, A.
Qadir, Riazuddin and H. Saleem, World Scientific, 2007.
\bibitem{dw} D. wiltshire, {\it Phys. Rev. Lett.} {\bf 99} (2007)
251101; {\it Phys. Rev.} {D 78} (2008) 084032; {\it Astrophys. J.}
{\bf 672} (2008) L91; ``Dark energy without dark energy",
arXiv:0712.3984 (Overview for Dark 2007 Proceedings).
\bibitem{caldwell} R.R. Caldwell,\textit{ Phys. Lett. }\textbf{B 545 }(2002) 23.
\bibitem{caldwell1} R.R. Caldwell et al,\textit{ Phys. Rev. Lett. }\textbf{91 }(2003)
071301.
\bibitem{qiu} T. Qiu et al,\textit{ Mod.Phys.Lett.}A23:2787-2798,2008.
\bibitem{pedro11} P.F.G. Diaz and C.L. Sigenza,\textit{ Phys. Lett. }\textbf{B 589}
(2004) 78.
\bibitem{fabris} K.A. Bronnikov and J.C. Fabris, \textit{Phys. Rev. Lett}. \textbf{96}
(2006) 251101
\bibitem{mar} M.P. Dabrowski, arXiv:gr-qc/0701057v1.
\bibitem{vinod} V. Johri,\textit{ Phys. Rev.}\textbf{D 70} (2004) 041303(R).
\bibitem{carroll} S.M. Carroll et al,\textit{ Phys. Rev. }\textbf{D 71 }(2005) 023525.
\bibitem{nojiri} S. Nojiri et al, \textit{ Phys.Rev.} \textbf{D71} (2005) 063004.
\bibitem{anton} A. Baushev, arXiv:0809.0235 [astro-ph].
\bibitem{carroll1} S. Carroll et al, \textit{Phys. Rev.} \textbf{D 68 }(2003) 023509.
\bibitem{pedro} P.F. Gonzalez-Diaz,\textit{ Phys.Rev.}\textbf{ D68 }(2003) 021303.
021102; arXiv:astro-ph/0505618v1; arXiv:gr-qc/0507119v1.
\bibitem{moruno} P.M. Moruno, \textit{Phys. Lett.} \textbf{B 659 }(2008) 40.
\bibitem{martin11} P.M. Moruno et al, \textit{Phys. Lett.}\textbf{ B 640} (2006) 117.
\bibitem{ness} S. Nesseris and L. Perivolaropoulos, \textit{Phys.Rev.} \textbf{D70 }(2004)
123529.
\bibitem{babichev} E. Babichev et al,\textit{ Phys. Rev. Lett.} \textbf{93 }(2004).
\bibitem{vikman} R. Akhoury et al, JHEP 0903:082,2009
\bibitem{jamil} M. Jamil et al,\textit{ Eur. Phys. J. }\textbf{C 58}(2008) 325.
\bibitem{martin} P.M. Moruno et al, arXiv:0803.2005v1 [gr-qc].
\bibitem{babichev1} E. Babichev et al,  \textit{Phys.Rev.}\textbf{D78 }(2008) 104027.
\bibitem{jose} J.A.J. Madrid and P.F.G. Diaz,
\textit{Grav.Cosmol.}\textbf{14 }(2008) 213.
\bibitem{nojiri1} S. Nojiri and S. Odintsov,\textit{ Phys.Rev.} \textbf{D70 }(2004)
103522.
\bibitem{gao} C. Gao et al, \textit{Phys. Rev.} \textbf{D 78 }(2008) 024008.
\bibitem{sun} C.Y. Sun,\textit{ Phys. Rev.} \textbf{D 78} (2008) 064060.
\bibitem{zhang} X. Zhang,  \textit{Eur. Phys. J.}\textbf{ C 60} (2009) 661.
\bibitem{haw} B.J. Carr and S.W. Hawking, \textit{Mon. Not. R. Astr. Soc.}
\textbf{168 }(1974) 399.
\bibitem{khlopov} M.Y. Khlopov et al, \textit{Mon. Not. R. Astr. Soc.} \textbf{215 }(1985) 575.
\bibitem{barrow1} J.D. Barrow and B.J. Carr, \textit{Phys. Rev.} \textbf{D 54 }(1996)
3920.
\bibitem{raf} R. Guedens et al,\textit{ Phys. Rev.}\textbf{ D66} (2002) 083509.
\bibitem{polarski} D. Polarski, \textit{Phys.Lett.} \textbf{B528} (2002) 193-198.
\bibitem{carr} B.J. Carr, arXiv: 0511743v1 [astro-ph].
\bibitem{carr1} B.J. Carr, arXiv:astro-ph/0102390.
\bibitem{pf} R. Penrose and R.M. Floyd, {\it Nature}, {\bf 229} (1971) 171.
\bibitem{swh} S.W. Hawking, {\it Phys. Rev. Lett.}, {\bf 26} (1971)
1344.
\bibitem{jacob} J.D. Bekenstein, \textit{Phys. Rev.} \textbf{D 7} (1973) 2333.
\bibitem{jacob1} J.D. Bekenstein,\textit{ Phys. Rev. }\textbf{D 9 }(1974) 3292.
\bibitem{jacob2} J.D. Bekenstein, \textit{Phys. Rev.} \textbf{D 10} (1975) 3077.
\bibitem{saf} S.A. Fulling, {\it Phys. Rev.} {\bf D 7} (1973) 2850.
\bibitem{hawking} S.W. Hawking, \textit{Commun. Math. Phys. }\textbf{43} (1975) 199.
\bibitem{hawk} S.W. Hawking, \textit{Nature} \textbf{248} (1974) 30.
\bibitem{page} D.N. Page, \textit{Phys. Rev. }\textbf{D 13} (1976) 198.
\bibitem{zel} Ya.B. Zeldovich and A.A. Starobinsky, \textit{JETP Lett.} \textbf{24}
(1976) 571.
\bibitem{novikov} I.D. Novikov et al, \textit{Astron. Astrophys.} \textbf{80 }(1979)
104.
\bibitem{barrow} J.D. Barrow, \textit{Mon. Not. R. Astr. Soc.} \textbf{192} (1980) 427.
\bibitem{bean} R. Bean and J. Magueijo, \textit{Phys. Rev. }\textbf{D 66} (2002) 063505.
\bibitem{book} \textit{Table of integrals, series and products} (Academic press,
1965).
\bibitem{siva} C. Sivaram, \textit{Gen. Relativ. Gravit.}\textbf{ 33 }(2001) 175.
\end{thebibliography}
\end{document}